# A Comment on the Magical Realizations of $W_3$

J. M. Figueroa-O'Farrill[†]

*Department of Physics, Queen Mary and Westfield College*
*Mile End Road, London E1 4NS, UK*

## ABSTRACT

In the process of investigating classical realizations of $W_3$ in terms of free bosons, Romans unveiled a relation to finite-dimensional Jordan algebras with a cubic norm. These algebras have been classified and consist of an infinite series (yielding the "generic" realizations) and four sporadic algebras associated to the real division algebras (which yield the "magical" realizations). The generic realizations were shown by Romans to quantize, who left the problem of the quantization of the magical realizations open. In later work, Mohammedi showed that the first two magical realizations did not survive quantization. In this note we close the problem by showing that neither do the other two magical realizations.

[†] e-mail: jmf@strings1.ph.qmw.ac.uk

## Introduction

One can usually learn a lot about an algebraic structure from the study of its concrete realizations. W-algebras are no exception and a lot of the effort in their study has been directed towards finding representations in terms of better understood algebras: free fields, parafermions, affine algebras,... Of these, the realizations in terms of free bosons are arguably the simplest and the most interesting, especially in view of the possible application to W-gravity and W-strings.

The earliest known such realization is the two-boson realization of Fateev and Zamolodchikov [1]; which is the analogue of the Coulomb gas realization for Virasoro. Because of the intrinsic nonlinearity of W-algebras, it is however not possible to now construct realizations in an arbitrary number of bosons by simply tensoring two-boson realizations together. Other ways must therefore be sought. A very general class of such realizations was found by Romans in [2], where he analyzed the problem of constructing realizations of the classical and quantum $W_3$-algebra in terms of free bosons. He found a classical realization associated to every Jordan algebra determined by a cubic norm. Explicitly, there is for every $n \in \mathbb{N}$ a "generic" realization in terms of $n$ bosons, whose corresponding Jordan algebra is essentially the Clifford algebra of (some real form of) $SO(n-1)$; and four "magic" solutions for $n = 5, 8, 14, 26$, whose corresponding Jordan algebras are the algebras (with respect to one half of the anticommutator) of $3 \times 3$ hermitian matrices over the real division algebras: the real numbers ($\mathbb{R}$), the complex numbers ($\mathbb{C}$), the quaternions ($\mathbb{H}$), and the octonions (or Cayley numbers) ($\mathbb{O}$) respectively. The generic series survives quantization (although for $n = 1$ it fixes the central charge to $c = -2$) and can be seen (for $n > 1$) to fit a general pattern of $W_3$ realizations whose ingredients are an arbitrary realization of the Virasoro algebra and a free boson commuting with it. The issue of the quantization of the magic realizations was left open and Mohammedi later showed [3] that the first two cases $n = 5$ and $n = 8$ do not quantize.

After this result there would be little reason to expect that the other two would quantize, were it not for the fact that there exists at least one example where a classical construction based on the real division algebras survives quantization only in the octonionic case [4]. Classical superstrings depend for their existence on the same $\gamma$-matrix identity which must be obeyed for supersymmetric Yang-Mills to exist. This identity is only satisfied in dimensions $D = 3, 4, 6, 10$ as a result of the idiosyncracies of the multiplication table of the real division algebras. Quantum mechanically, however, only the $D = 10$ string survives. Unfortunately for us, this happy circumstance does not repeat itself in our problem. We will see that none of the magical $W_3$ realizations survives quantization.

This note is organized as follows. We will set the context and the notation by briefly reviewing the results of Romans and we will immediately focus on the magical solutions and their relation with the special Jordan algebras. We then proceed to a case by case analysis of the four magical solutions, giving in each case the form of the classical realizations. Finally we comment on their failure to quantize.

Realizations of $\mathsf{W}_3$

Let us summarize some of the results of [2] concerning the realizations of $\mathsf{W}_3$ in terms of $D$ scalar bosons $\varphi_i(z)$ with OPE

$$\varphi_i(z)\varphi_j(w) = -g_{ij}\log(z-w) + \cdots . \qquad (1)$$

We will only work with their currents $J_j(z) = i\partial\varphi_j(z)$. The expressions for $T$ and $W$ in terms of these bosons have the form

$$\begin{aligned} T &= \tfrac{1}{2}g^{ij}J_iJ_j + a^i\partial J_i \\ W &= \tfrac{1}{6}d^{ijk}J_iJ_jJ_k + e^{ij}J_i\partial J_j + e^i\partial^2 J_i , \end{aligned} \qquad (2)$$

where $g^{ij}$ is the inverse of the $g_{ij}$ appearing in (1). $T$ obeys a Virasoro algebra with central charge $c = D - 12a^2$ with $a^2 \equiv g_{ij}a^ia^j$; and $W$ is a conformal primary of weight 3 provided that the following equations are satisfied:

$$e^{(ij)} = 2a_k d^{ijk} \qquad (3)$$
$$e^i = \tfrac{1}{3}a_j e^{ji} \qquad (4)$$
$$d^{ij}{}_j = 12 e^{ij}a_j + 4 a_j e^{ji} \qquad (5)$$

from where we can eliminate $e^i$ and the symmetric part of $e^{ij}$. Let us therefore introduce $F^{ij} = e^{[ij]} = \tfrac{1}{2}(e^{ij} - e^{ji})$. Then we can use (3) to rewrite (5) in the following form:

$$d^{ij}{}_j = 16 F^{ij}a_j + 4 d^{ijk}a_j a_k . \qquad (6)$$

The conditions for $W$ to close into the $\mathsf{W}_3$ algebra

$$\begin{aligned} W(z)W(w) &= \frac{\tfrac{c}{3}}{(z-w)^6} + \frac{2T(w)}{(z-w)^4} + \frac{\partial T(w)}{(z-w)^3} + \frac{\mu\Lambda(w) + \tfrac{3}{10}\partial^2 T(w)}{(z-w)^2} \\ &\quad + \frac{\tfrac{1}{2}\mu\partial\Lambda(w) + \tfrac{1}{15}\partial^3 T(w)}{z-w} , \end{aligned} \qquad (7)$$

where $\mu = \frac{16}{5c+22}$ and $\Lambda = (TT) - \tfrac{3}{10}\partial^2 T$, can be shown to reduce to

$$d^{(ij}{}_m d^{kl)m} = 2\mu g^{(ij}g^{kl)} \qquad (8)$$

$$d^l{}_{jk}F_{li} + d^l{}_{i(j}e_{k)l} = \mu a_i g_{jk} \qquad (9)$$
$$d_i{}^{kl}d_{jkl} + 12 d_{ijk}e^k - 12 e_{ik}e^k{}_j = 2 g_{ij} . \qquad (10)$$

One recognizes in equation (8) the condition for the existence of a classical realization of $\mathsf{W}_3$:

$$\begin{aligned} T^{\text{class}} &= \tfrac{1}{2}g^{ij}J_iJ_j \\ W^{\text{class}} &= \tfrac{1}{6}d^{ijk}J_iJ_jJ_k , \end{aligned} \qquad (11)$$

where we now take the Poisson bracket or equivalently the terms containing exactly one derivative in the simple pole of the OPE.

It is precisely equation (8) which connects us with the theory of Jordan algebras possessing a cubic norm. Romans exploited this relation to classify the solutions to (8). There are two classes of solutions. First of all there is an infinite series for $D = 1, 2, \ldots$ and four sporadic cases for $D = 5, 8, 14, 26$. The infinite series yields the "generic" realizations which shall not concern us here. Romans showed that they did survive quantization; that is, that $F^{ij}$ can be found such that all equations (6), (9), and (10) are satisfied, yielding realizations of quantum $\mathsf{W}_3$ for arbitrary of the cetral charge $c$ except for the case $D = 1$ which forces $c = -2$. The sporadic cases yield the "magical" realizations, which shall be the focus of this paper. Romans left open the problem of the quantization of the magical realizations. The issue was reconsidered in a later paper by Mohammedi [3], who showed that the cases $D = 5$ and $D = 8$ did not survive quantization. We will recap this work in what follows for completeness and to present the problem fully in a Jordan-algebraic approach.

As we shall see in the next section, the magical realizations all share the properties that the $d_{ijk}$ are traceless: $g^{ij}d_{ijk} = 0$. It is not hard to show that for $D \neq 2$, this condition implies that $a^i = e^i = 0$ and that $e^{ij} = F^{ij}$. In this case, equation (10) reduces to $F_{[i}{}^\ell d_{j]k\ell} = 0$ and tracing (8) we can turn (9) into

$$F^i{}_k F^k{}_j = -\frac{\tfrac{1}{2}D - 1}{5D + 22}\delta^i_k . \qquad (12)$$

We will find it convenient to rescale $F_{ij}$ and $d_{ijk}$ to hide the $D$-dependence away in the expression for $W$. Hence we have

$$W = \tfrac{1}{6}\alpha d^{ijk}J_iJ_jJ_k + \beta F^{ij}J_i\partial J_j , \qquad (13)$$

where $\alpha^2 = \frac{96}{5D+22}$ and $\beta^2 = \frac{\tfrac{1}{2}D-1}{5D+22}$. With this choice of normalizations, the magical realizations quantize provided that the following equations are satisfied:

$$d_{(ij}{}^r d_{kl)r} = \tfrac{1}{3}g_{(ij}g_{kl)} \qquad (14)$$
$$F_{[i}{}^\ell d_{j]k\ell} = 0 , \qquad (15)$$

and

$$F^i{}_k F^k{}_j = -\delta^i_j \ . \tag{16}$$

In what follows we will discuss how the special Jordan algebras yield solutions to (14). We will describe the solutions in detail and we will show that they fail to quantize. Notice, however, that without doing any work we can already discard the case $D = 5$ since equation (16) says that $F^k{}_i$ is a complex structure, whence $D$ must be even.

Special Jordan algebras

We now describe the magical realizations of classical $\mathsf{W}_3$ from the point of view of special Jordan algebras.

Let $\mathbb{A}$ denote one of the four real division algebras: $\mathbb{R}$, $\mathbb{C}$, $\mathbb{H}$, $\mathbb{O}$. Each algebra comes equipped with a conjugation $\alpha \mapsto \bar{\alpha}$ which leaves invariant only the real subalgebra generated by the identity. We now consider the vector space $\mathbb{J}_3(\mathbb{A})$ of $3 \times 3$ hermitian matrices with entries in $\mathbb{A}$:

$$\mathbb{J}_3(\mathbb{A}) \ni x = \begin{pmatrix} a_1 & \alpha_1 & \alpha_2 \\ \bar\alpha_1 & a_2 & \alpha_3 \\ \bar\alpha_2 & \bar\alpha_3 & a_3 \end{pmatrix} \ , \tag{17}$$

with $\alpha_i \in \mathbb{A}$ and $a_i \in \mathbb{R}$. On $\mathbb{J}_3(\mathbb{A})$ we can define a product $\circ$ by one half the anticommutator: $x \circ y = \tfrac{1}{2}(xy + yx)$. This product is commutative but non-associative; although it obeys the defining relation for a Jordan algebra:

$$x \circ (y \circ x^2) = (x \circ y) \circ x^2 \ . \tag{18}$$

The matrix trace defines a linear map $\operatorname{Tr} : \mathbb{J}_3(\mathbb{A}) \to \mathbb{R}$, which induces a split

$$\mathbb{J}_3(\mathbb{A}) = \ker \operatorname{Tr} \oplus \mathbb{R} \cdot \mathbf{1} \ , \tag{19}$$

with $\mathbf{1}$ the identity matrix. This decomposition is orthogonal relative to the bilinear form

$$\langle x, y \rangle \equiv \tfrac{1}{2} \operatorname{Tr}(x \circ y) \ . \tag{20}$$

Choose a basis $\{e_i\}$ for the traceless elements. Then we have

$$e_i \circ e_j = \tfrac{2}{3} g_{ij} \mathbf{1} + d_{ij}{}^k e_k \tag{21}$$

for some $d_{ij}{}^k$ and $g_{ij}$ which will turn out the ones appearing in the expressions (2) (after taking into account the normalization (13)). We can recover the $g$- and $d$-symbols as follows. It follows that $g_{ij} = \langle e_i, e_j \rangle$ and that

$$\langle e_i \circ e_j, e_k \rangle = d_{ij}{}^l g_{lk} \equiv d_{ijk} \ . \tag{22}$$

The coefficients $g_{ij}$ and $d_{ijk}$ are the restriction to the traceless elements of the Jordan algebra of a quadratic and cubic norms defined respectively by (20) and by

$$C(x, y, z) \equiv \langle x \times y, z \rangle \ , \tag{23}$$

where the Freudenthal product is defined by

$$x \times y \equiv x \circ y - \tfrac{1}{2} x \operatorname{Tr} y - \tfrac{1}{2} y \operatorname{Tr} x + \tfrac{1}{2}(\operatorname{Tr} x \operatorname{Tr} y - \operatorname{Tr} x \circ y)\mathbf{1} \ . \tag{24}$$

For the case of $\mathbb{A} = \mathbb{C}$, (21) is a famous identity satisfied by the Gell-Mann $\lambda$-matrices and where $d_{ijk}$ are the $SU(3)$ $d$-symbols. These symbols are well-known to be totally symmetric and traceless: $g^{ij} d_{ijk} = 0$. This continues to be the case for the other three division algebras. Moreover one can show that in this normalization, the $d$-symbols satisfy (14), whence they define a classical realization of the $\mathsf{W}_3$ algebra. Below we list the $d$-symbols for the special Jordan algebras. Since we already proved that the realization associated to $\mathbb{J}_3(\mathbb{R})$ does not quantize we focus on the remaining three cases. The calculations of the $d$-symbols were performed in Mathematica®.

Case I: $\mathbb{J}_3(\mathbb{C})$

We choose as basis for the Jordan algebra the identity matrix together with the following eight hermitian traceless matrices:

$$e_1 = \begin{pmatrix} 0 & 1 & 0 \\ 1 & 0 & 0 \\ 0 & 0 & 0 \end{pmatrix} \quad e_2 = \begin{pmatrix} 0 & 0 & 1 \\ 0 & 0 & 0 \\ 1 & 0 & 0 \end{pmatrix} \quad e_3 = \begin{pmatrix} 0 & 0 & 0 \\ 0 & 0 & 1 \\ 0 & 1 & 0 \end{pmatrix} \ ,$$

$$e_4 = \begin{pmatrix} 0 & i & 0 \\ -i & 0 & 0 \\ 0 & 0 & 0 \end{pmatrix} \quad e_5 = \begin{pmatrix} 0 & 0 & i \\ 0 & 0 & 0 \\ -i & 0 & 0 \end{pmatrix} \quad e_6 = \begin{pmatrix} 0 & 0 & 0 \\ 0 & 0 & i \\ 0 & -i & 0 \end{pmatrix} \ ,$$

and

$$e_7 = \begin{pmatrix} 1 & 0 & 0 \\ 0 & -1 & 0 \\ 0 & 0 & 0 \end{pmatrix} \quad e_8 = \frac{1}{\sqrt{3}} \begin{pmatrix} 1 & 0 & 0 \\ 0 & 1 & 0 \\ 0 & 0 & -2 \end{pmatrix} \ . \tag{25}$$

With this choice of basis, it is easy to see that $g_{ij} = \delta_{ij}$ and that the nonzero

$d_{ijk}$ are given by $d_{ijk} = \frac{1}{4} \mathrm{Tr}(e_i e_j e_k + e_j e_i e_k)$; that is,

$$d_{118} = d_{448} = d_{778} = -d_{888} = \frac{1}{\sqrt{3}},$$
$$d_{228} = d_{338} = d_{558} = d_{668} = -\frac{1}{2\sqrt{3}}, \quad (26)$$
$$d_{123} = d_{156} = d_{227} = -d_{246} = -d_{337} = d_{345} = d_{557} = -d_{667} = \tfrac{1}{2}.$$

Notice parenthetically, that $e_1$, $e_2$, $e_3$, together with $e_7$ and $e_8$ span a Jordan subalgebra isomorphic to $\mathbb{J}_3(\mathbb{R})$. We can thus read its $d$-symbols from these if need be.

### Case II: $\mathbb{J}_3(\mathbb{H})$

Let $\{q_i\}$ denote a basis for the quaternions with $q_0 = 1$, $q_1 = i$, $q_2 = j$, and $q_3 = k$ with $i, j, k$ the usual quaternion imaginary units. As their names suggest, $q_0$ is real: $\bar{q}_0 = q_0$, whereas the other units obey $\bar{q}_i = -q_i$ for $i = 1, 2, 3$. As basis for the traceless subspace of the Jordan algebra we now choose the following 14 hermitian matrices:

$$e_{3i+1} = \begin{pmatrix} 0 & q_i & 0 \\ \bar{q}_i & 0 & 0 \\ 0 & 0 & 0 \end{pmatrix} \quad e_{3i+2} = \begin{pmatrix} 0 & 0 & q_i \\ 0 & 0 & 0 \\ \bar{q}_i & 0 & 0 \end{pmatrix} \quad e_{3i+3} = \begin{pmatrix} 0 & 0 & 0 \\ 0 & 0 & q_i \\ 0 & \bar{q}_i & 0 \end{pmatrix},$$

for $i = 0, 1, 2, 3$, and

$$e_{13} = \begin{pmatrix} 1 & 0 & 0 \\ 0 & -1 & 0 \\ 0 & 0 & 0 \end{pmatrix} \quad e_{14} = \frac{1}{\sqrt{3}} \begin{pmatrix} 1 & 0 & 0 \\ 0 & 1 & 0 \\ 0 & 0 & -2 \end{pmatrix}. \quad (27)$$

With this choice of basis, it is easy to see that again $g_{ij} = \delta_{ij}$ and that the nonzero $d_{ijk}$ are given by

$$d_{1,1,14} = d_{4,4,14} = d_{7,7,14} = d_{10,10,14} = d_{13,13,14} = -d_{14,14,14} = \frac{1}{\sqrt{3}},$$

$$d_{5,5,14} = d_{6,6,14} = d_{2,2,14} = d_{3,3,14} = d_{8,8,14} = d_{9,9,14} = d_{11,11,14}$$
$$= d_{12,12,14} = -\frac{1}{2\sqrt{3}},$$
$$d_{1,2,3} = d_{1,5,6} = d_{1,8,9} = d_{1,11,12} = d_{2,2,13} = d_{6,8,10} = -d_{2,4,6} = -d_{2,7,9}$$
$$= -d_{2,10,12} = -d_{3,3,13} = d_{3,4,5} = d_{3,7,8} = d_{3,10,11} = -d_{4,8,12} = d_{4,9,11}$$
$$= d_{5,5,13} = d_{5,7,12} = -d_{5,9,10} = -d_{6,6,13} = -d_{6,7,11} = d_{8,8,13} = -d_{9,9,13}$$
$$= d_{11,11,13} = -d_{12,12,13} = \tfrac{1}{2}. \quad (28)$$

Notice that $e_{1-6}$, together with $e_{13}$ and $e_{14}$ span a Jordan subalgebra isomorphic to $\mathbb{J}_3(\mathbb{C})$.

### Case III: $\mathbb{J}_3(\mathbb{O})$

In order to describe this algebra it is necessary to describe the octonion algebra itself. We prefer to think of octonions as pairs of quaternions: $\mathbb{O} \cong \mathbb{H} \oplus \mathbb{H}$. On this space we define the following multiplication

$$(a,b) \circ (c,d) = (ac - \bar{d}b, da + b\bar{c}), \quad (29)$$

and the following conjugation

$$\overline{(a,b)} = (\bar{a}, -b). \quad (30)$$

A basis for $\mathbb{O}$ is given by the identity $(1,0)$ together with $x_1$ through $x_7$ defined by: $(i,0)$, $(j,0)$, $(k,0)$, $(0,1)$, $(0,i)$, $(0,j)$, and $(0,k)$. They obey $x_p \circ x_p = -1$, $x_p \circ x_q = -x_q \circ x_p$ for $p \neq q$, and $\bar{x}_p = -x_p$. The multiplication table can be pictured as follows:

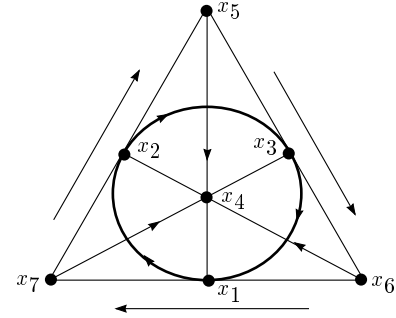

where, if following any line in the directions shown by the arrows, one encounters $x_p \to x_q \to x_r$ in this order, then $x_p \circ x_q = x_r$ and $x_q \circ x_r = x_p$. We must supplement this multiplication table by the fact that $x_0 \equiv (1,0)$ is the identity of the algebra. It is easy to check that the multiplication is neither commutative nor associative.

The traceless elements of the exceptional Jordan algebra $\mathbb{J}_3(\mathbb{O})$ are spanned by the following 26 matrices:

$$e_{3i+1} = \begin{pmatrix} 0 & x_i & 0 \\ \bar{x}_i & 0 & 0 \\ 0 & 0 & 0 \end{pmatrix} \quad e_{3i+2} = \begin{pmatrix} 0 & 0 & x_i \\ 0 & 0 & 0 \\ \bar{x}_i & 0 & 0 \end{pmatrix} \quad e_{3i+3} = \begin{pmatrix} 0 & 0 & 0 \\ 0 & 0 & x_i \\ 0 & \bar{x}_i & 0 \end{pmatrix},$$

for $i = 0, 1, \ldots, 7$, and

$$e_{25} = \begin{pmatrix} 1 & 0 & 0 \\ 0 & -1 & 0 \\ 0 & 0 & 0 \end{pmatrix} \quad e_{26} = \frac{1}{\sqrt{3}} \begin{pmatrix} 1 & 0 & 0 \\ 0 & 1 & 0 \\ 0 & 0 & -2 \end{pmatrix}. \quad (31)$$

With this choice of basis, it is easy to see that again $g_{ij} = \delta_{ij}$. The nonzero

$d_{ijk}$ are given this time by

$$d_{1,1,26} = d_{4,4,26} = d_{7,7,26} = d_{10,10,26} = d_{13,13,26} = d_{16,16,26} = d_{19,19,26}$$
$$= d_{22,22,26} = d_{25,25,26} = -d_{26,26,26} = \frac{1}{\sqrt{3}} ,$$
$$d_{2,2,26} = d_{3,3,26} = d_{5,5,26} = d_{6,6,26} = d_{8,8,26} = d_{9,9,26} = d_{11,11,26}$$
$$= d_{12,12,26} = d_{14,14,26} = d_{15,15,26} = d_{17,17,26} = d_{18,18,26} = d_{20,20,26} = d_{21,21,26}$$
$$= d_{23,23,26} = d_{24,24,26} = -\frac{1}{2\sqrt{3}} ,$$
$$d_{1,2,3} = d_{1,5,6} = d_{1,8,9} = d_{1,11,12} = d_{1,14,15} = d_{1,17,18} = d_{1,20,21}$$
$$= d_{1,23,24} = d_{2,2,25} = -d_{2,4,6} = -d_{2,7,9} = -d_{2,10,12} = -d_{2,13,15} = -d_{2,16,18}$$
$$= -d_{2,19,21} = -d_{2,22,24} = -d_{3,3,25} = d_{3,4,5} = d_{3,7,8} = d_{3,10,11} = d_{3,13,14}$$
$$= d_{3,16,17} = d_{3,19,20} = d_{3,22,23} = -d_{4,8,12} = d_{4,9,11} = -d_{4,14,18} = d_{4,15,17}$$
$$= d_{4,20,24} = -d_{4,21,23} = d_{5,5,25} = d_{5,7,12} = -d_{5,9,10} = d_{5,13,18} = -d_{5,15,16}$$
$$= -d_{5,19,24} = d_{5,21,22} = -d_{6,6,25} = -d_{6,7,11} = d_{6,8,10} = -d_{6,13,17} = d_{6,14,16}$$
$$= d_{6,19,23} = -d_{6,20,22} = -d_{7,14,21} = d_{7,15,20} = -d_{7,17,24} = d_{7,18,23} = d_{8,8,25}$$
$$= d_{8,13,21} = -d_{8,15,19} = d_{8,16,24} = -d_{8,18,22} = -d_{9,9,25} = -d_{9,13,20} = d_{9,14,19}$$
$$= -d_{9,16,23} = d_{9,17,22} = -d_{10,14,24} = d_{10,15,23} = d_{10,17,21} = -d_{10,18,20}$$
$$= d_{11,11,25} = d_{11,13,24} = -d_{11,15,22} = -d_{11,16,21} = d_{11,18,19} = -d_{12,12,25}$$
$$= -d_{12,13,23} = d_{12,14,22} = d_{12,16,20} = -d_{12,17,19} = d_{14,14,25} = -d_{15,15,25}$$
$$= d_{17,17,25} = -d_{18,18,25} = d_{20,20,25} = -d_{21,21,25} = d_{23,23,25} = -d_{24,24,25} = \tfrac{1}{2} .$$
(32)

<u>The failure of quantization</u>

The quantization of the classical realizations described above reduces to finding $F^{ij}$ satisfying equations (15) and (16). Unfortunately I have not been able to find an elegant proof of the inconsistency of these equations, and have had to resort to their brute force solution. In all cases, as had been found already for the case $\mathbb{A} = \mathbb{C}$ in [3], the unique solution to the linear equations (15) is $F^{ij} = 0$ which is incompatible with the quadratic equation (16). The failure of the linear equation to give any nontrivial solution certainly hints at the fact that a more global approach to this problem must exist that yields this result at one fell swoop.

ACKNOWLEDGEMENTS

It is a pleasure to thank Noureddine Mohammedi, Leonardo Palacios, Eduardo Ramos, and Sonia Stanciu for helpful discussions, and in particular Chris Hull for reminding me of [4] and for the subsequent encouragement.